\documentclass[10pt]{article}

\usepackage[superscript,sort]{cite}

\usepackage{ifthen,ifpdf}
\ifpdf
	\usepackage[pdftex]{hyperref}	 \hypersetup{colorlinks=true,linkcolor=black,citecolor=black,urlcolor=blue,backref=page,bookmarks=true,breaklinks=true,plainpages=false }
	\usepackage[pdftex]{graphicx}
	\usepackage[usenames,pdftex]{color}
\else
	\usepackage[colorlinks=true,breaklinks=true,plainpages=false]{hyperref}
	\usepackage{graphicx}
	\usepackage[usenames]{color}
\fi

\usepackage{amsthm,amscd,amsxtra,amsfonts,amsmath,amssymb,multirow}
\usepackage{wrapfig}
\usepackage[footnotesize]{caption}
\usepackage[tiny,compact]{titlesec}
\usepackage[textwidth=0.8in,textsize=footnotesize]{todonotes}

\usepackage{longtable}
\usepackage{multirow}
\usepackage{algorithm}
\usepackage[noend]{algpseudocode}

\usepackage{helvet}

\usepackage[letterpaper,top=0.5in,bottom=0.5in,left=0.5in,right=0.5in]{geometry}

\titlespacing*{\section}{0pt}{*0}{*0}
\titlespacing*{\subsection}{0pt}{*0}{*0}
\titlespacing*{\subsubsection}{0pt}{*0}{*0}
\titlespacing{\paragraph}{0pt}{*0}{*1}
\definecolor{MyPurple}{rgb}{1,0,1}

\begin{document}

%\pagenumbering{roman}
%\begin{verbatim}

\title{Automatic parametrization of implicit solvent models for the blind prediction of solvation free energies}

\author{Bao Wang$^{1}$, Zhixiong Zhao$^{1}$, and Guo-Wei Wei$^{1,2,3}$
\footnote{
Correspondences should be addressed  to Guo-Wei Wei. E-mail: wei@math.msu.edu}  \\
$^1$Department of Mathematics \\
Michigan State University, MI 48824, USA\\
$^2$Department of Electrical and Computer Engineering \\
Michigan State University, MI 48824, USA \\
$^3$Department of Biochemistry and Molecular Biology\\
Michigan State University, MI 48824, USA
}

\date{\today}

\maketitle

\begin{abstract}
In this work, a systematic protocol is proposed to automatically parametrize implicit solvent models with polar and nonpolar components. The proposed protocol utilizes the classical Poisson model or the Kohn-Sham density functional theory (KSDFT) based polarizable Poisson model for modeling   polar solvation free energies. For the nonpolar component, either the standard model of surface area, molecular volume, and van der Waals interactions, or a model with atomic surface areas and molecular volume is employed. Based on the assumption that similar molecules have similar parametrizations, we develop scoring and ranking algorithms to classify solute molecules.  Four sets of radius parameters are combined with four sets of charge force fields to arrive at a total of 16 different parametrizations for the Poisson model. A large database with 668 experimental data is utilized to validate the proposed protocol.  The lowest leave-one-out root mean square (RMS) error for the database is 1.33k cal/mol. Additionally, five subsets of the database, i.e., SAMPL0-SAMPL4,     are employed to further demonstrate that the proposed protocol offers some of the best solvation predictions. The optimal RMS errors are 0.93, 2.82, 1.90, 0.78, and  1.03 kcal/mol, respectively for SAMPL0, SAMPL1, SAMPL2, SAMPL3, and SAMPL4 test sets. These results are some of the best, to our best knowledge.

\end{abstract}


\newpage

 {\setcounter{tocdepth}{4} \tableofcontents}

\newpage
\section{Introduction}
Solvation in which separated solvent and solute molecules are combined to form a solution, is an elementary physical process in nature that provides a foundation for more complicated chemical and biological processes, such as ion channel permeation, protein-ligand binding,  electron transfer, signal transduction, DNA specification, transcription, post-transcription modification, gene expression, protein synthesis, etc. Therefore, the understanding of solvation is a  prerequisite for the quantitative study of other more complex chemical and biological processes and is of paramount importance to chemistry, physics and biology  \cite{SolvationTruhlarAndCramer:1994,Daudel:1973, Reichardt:1990,SolvationTruhlarAndCramer:1994}. The most basic and reliable experimental observation of solvation is the solvation free energy, which  measures the free energy change  in the solvation process. As a result, one of the most   important tasks in the solvation modeling and computation is the prediction of solvation free energies, which has recently drawn much attention in computational biophysics and chemistry \cite{Daudel:1973,Kreevoy:1986, Reichardt:1990,SolvationTruhlarAndCramer:1994,Davis:1990a, Warshel:1998BW}.

%As a result, solvation modeling and computation are essentially about solvation free energy analysis and prediction, which have recently drawn much attention in chemical and biological sciences \cite{Daudel:1973,Kreevoy:1986, Reichardt:1990,SolvationTruhlarAndCramer:1994,Davis:1990a, Warshel:1998BW}.

%Current solvation models, Born-MD
A large variety of solvation models has been developed in the past few decades for
solvation analysis and solvation free energy prediction. In general, these models can
be classified into either knowledge based models or physics based models. Knowledge
based models usually utilize a large amount of available data of solvation free
energies to train statistical models for the solvation free energy prediction.
One example of this type of models makes use of solvent accessible surface areas,
solvent excluded surface areas or calibrated atomic surface areas for solvation free
energy predictions \cite{Kollman:2001Solvation}. Knowledge based models can provide relatively
accurate predictions provided that enough experimental data are available. On the
other hand, physics based solvation models are based on fundamental laws of physics.
These models can be further classified into three major classes. One of them is explicit solvent models in which both solvent and
solute molecules are described at the atomic or electronic level. This class of models is
accurate in general, but is usually computationally expensive \cite{Savelyev:2007, Lin:2002, Koehl:2006}.
Another class of models is integral equation based solvation theories \cite{Hirata:2003, Fedorov:2010, Borgis:2005},
in which the solute molecule is  still modeled at the atomic level, while the solvent is modeled by
statistical mechanics, such as liquid density functional theory, Ornstein-Zernike equation with hypernetted-chain, Percus-Yevick
equation or other specified closure. Integral equation based solvation models reduce the number of
degrees of freedom dramatically compared to explicit solvent models. A major feature of integral equation
approaches is that these  methods are able to provide a good approximation of solvent microstructures near the
solvent-solute  interface \cite{Giambasu:2014}. % However, the deficient of this model is lack the knowledge for selection of the unified approximation of bridge functional.
The other class of the physics based solvation models is implicit solvent
models \cite{Orozco:2000, Truhlar:1999SolvationChemRev, Roux:1999, Prabhu:2004, Gallicchio:2004, Baker:2005,Geng:2007a, Wei:2009, ZhanChen:2010a},
in which the solute molecule can be modeled at either  the atomic or the quantum mechanical level, while the solvent is
described as a dielectric continuum. %Implicit solvent models further reduce  the number of degrees of freedom compared to the integral equation based solvation models.

%Implicit solvent based solvation models, from Born, Kirkwood,  Generalized Born, PCM to PB
Compared to  other methods, implicit solvent models have advantages of involving a
small number of degrees of freedom, highly accurate modeling of strong and
long-range electrostatic interactions that often dominate solvation phenomena,
and convenient treatments of solvent-solute electronic polarization effects  \cite{Truhlar:1999SolvationChemRev}.
In implicit solvent models, the solvation free energy is typically divided into polar and
nonpolar components. The polar solvation free energy is due to electrostatic
interactions and can be modeled by a number of approaches, including Born
dielectric sphere model, Kirkwood model \cite{Kirkwood:1934}, generalized Born (GB) model \cite{Jayaram:1998, Grant:2007, Gohlke:2004, Feig:2004c, Dominy:1999, Bashford:2000},
polarizable continuum model (PCM) \cite{Jacopo:2005, PCM:1994, PCM:1997} and
Poisson-Boltzmann (PB) theory \cite{Chung:2003, Sharp:1990,  Rocchia:2001, Honig:1995a,Gilson:1993}.
Among these approaches, Born dielectric sphere model which treats the solute molecule
as the dielectric sphere with a centrally located atomic charge is the simplest one.
The Kirkwood model gives analytical expressions for electrostatic potentials in a
spherical solute-solvent system with multiple charges inside the solute molecule \cite{Kirkwood:1934}.
The GB model \cite{Jayaram:1998, Grant:2007, Gohlke:2004, Feig:2004c, Dominy:1999, Bashford:2000}
is able to deal with arbitrarily shaped molecules and account for the impact of each
point charge by its effective distance (i.e., the Born radius) from the solvent-solute
interface \cite{Tan:2006c,Truhlar:2008Solvation}.
 %it is a currently widely used implicit solvent model. Typical two types of software suites based on the GB implicit solvent theory are GBSA implicit solvent molecular dynamics simulation package \cite{Tan:2006c}, and the other one are the SM.x solvation energy prediction softwares \cite{Truhlar:1999SolvationChemRev, %Truhlar:2006Solvation, Truhlar:2001Solvation2, Truhlar:1999Solvationext, Truhlar:2001Solvation, Truhlar:2008Solvation}.
The GB model is relatively faster, while depending on other methods, such as the PB theory,
for its parametrization. The PCM describes solute electrons by using quantum mechanics
so that solvation induced polarizations can be accounted in an iterative procedure. A few different versions of PCM models have been reported,  including  dielectric PCM, integral equation formalism PCM,
 and the variational PCM models \cite{Jacopo:2005, PCM:1994, PCM:1997}. PCM is relatively computationally expensive due
 to its quantum mechanical charge calculation while its accuracy is bounded  by the PB type of treatment of electrostatic potential
 as well as the quality of solvent-solute interfaces. The PB theory can be derived from more fundamental Maxwell's
 equations \cite{Chung:2003, Sharp:1990,  Rocchia:2001, Honig:1995a,Geng:2007a}. It is more accurate than the GB model
 and computationally more efficient than the PCM. The PB can be easily coupled with the quantum mechanics for a more
 accurate description of the solute charge density  \cite{MLWang:2006, MLWang:2007,ZhanChen:2011a}. It has become one of the
 most popular solvation models due to its relatively low computational cost and high modeling accuracy for biomolecules.
%which models the ion distribution in the solvent based Boltzmann distribution assumption, the PB model is a widely accepted efficient solvation model in the molecular biological science \cite{Chung:2003, Sharp:1990,  Rocchia:2001, Honig:1995a}. The PB can be easily coupled with the quantum mechanics for a more accurate description of the solute molecules \cite{MLWang:2006, MLWang:2007}, which is also widely used for the implicit molecular dynamics simulation and protein ligand binding free energy prediction\cite{Geng:2011, Weinzinger:2005, Tan:2006c, Swanson:2004, Lin:2003, Gouda:2003, Fogolari:2003}.

The nonpolar solvation component has been modeled by a number of terms. A popular approach, based on the scaled-particle theory (SPT) for nonpolar solutes in aqueous solutions \cite{Stillinger:1973,Pierotti:1976}, is to use a solvent-accessible surface area (SASA) term \cite{Swanson:2004,massova2000combined}. It was shown that a solvent-accessible volume (SAV) term is relevant in large length scale regimes \cite{Lum:1999,huang2000temperature}. Recent studies indicates that SASA based solvation models may not describe van der Waals (vdW) interactions near solvent-solute interface \cite{Gallicchio:2002,Gallicchio:2004,choudhury2005mechanism,Wagoner:2006}. A combination of surface area, surface enclosed volume, and vdW potential has been shown to provide accurate nonpolar solvation predictions  \cite{ZhanChen:2012,BaoWang:2015a}.

The polar and nonpolar components are typically decoupled in traditional implicit solvent models. Recently, new efforts have been made to couple polar and nonpolar components in the literature \cite{Dzubiella:2006,Wei:2009, ZhanChen:2010a}.  Differential geometry based solvation model makes use of the differential geometry theory of surfaces to dynamically couple polar and nonploar models by surface evolution, which is driven by free energy optimization \cite{Wei:2009, ZhanChen:2011a, ZhanChen:2010a, ZhanChen:2010b}. Coupled with an optimization procedure, this model was shown to deliver some of the best nonpolar solvent free energy analysis \cite{ZhanChen:2012}. By applying constrained optimization to nonpolar parameter selection, this model provides state-of-the-art solvation free energy fitting and cross validation results for a large number of solute molecules \cite{BaoWang:2015a}.

%Sampl solvation test
It is important to validate solvation models by experimental data. SAMPLx blind solvation prediction project, aimed at testing protein ligand binding models, also provides benchmark tests of the performance of the solvation models. Many different solvation approaches have been tested by the SAMPLx challenging molecules \cite{Nicholls:2008solvation, David:2014solvation, Guthrie:2014solvation, Klamt:2014solvation, David:2012solvation, David:2009solvation,  David:2006JCTCsolvation, David:2008JPCBsolvation, Wujianzhong:2014, SAMPL1:2009, SAMPL:2010, SAMPL3:2012}. Implicit solvent models are among the most competitive approaches in SAMPLx tests.

%Data set introduction
The objective of the present work is to develop an automatic  protocol to parametrize implicit solvent models for the {\it blind} solvation free energy prediction. In our approach, we utilize PB theories for computing electrostatic solvation free energies. Both a point charge based PB model and a KSDFT based polarizable PB model are developed in the present work. For the KSDFT based PB model, an iterative process is developed to take care of the solvent polarization and solute response \cite{ZhanChen:2011a}. In our nonpolar models, we employ either a collection of surface area, molecular volume and van der Waals interactions or a combination of atomic surface areas and molecular volume. To determine parameters in our nonpolar models,  we assume that similar molecules admit the same set of nonpolar parameters. In this work, we develop two parametrization algorithms. One of them is based on  a functional group scoring, and the other is based on a nearest neighbor search. Both algorithms utilize machine learning methods to optimize parameters.
A database contains the solvation free energies in aqueous solvent for 668 solute molecules  is collected from the literature \cite{Cabani:1981,Mobley:2014} to validate the proposed parametrization protocol. It is found that the present approach provides some of the best blind predictions of solvation free energies.

%Organization of this paper
The rest of this paper is organized as follows. In section \ref{ModelMethod}, we present our solvation models. Both polar and nonpolar models are discussed.  Sections \ref{NonpolarEnergyModel1} and \ref{NonpolarEnergyModel2} present two nonpolar parametrization algorithms, namely, functional group scoring and nearest neighbor approaches. An algorithm for unified blind solvation free energy prediction is developed in Section \ref{SolvationModel}. The blind prediction of a large number  of solvation free energies is carried out  in Section \ref{NumericalResults}. The leave-one-out test of 668 molecules is employed to validate the proposed protocol. Finally, the accuracy and robustness of the proposed protocol is demonstrated by five SAMPL test sets. This paper ends with a conclusion.

\section{Models and algorithms}\label{ModelMethod}

\subsection{Solvation models}

As an implicit approach,  our solvation model has polar and nonpolar components
\begin{equation}
\label{TotalModels}
\Delta G = \Delta G^{\rm p} + G^{\rm np},
\end{equation}
where  $\Delta G $ is the total solvation free energy, $\Delta G^{\rm p}$ is the polar or electrostatic solvation free energy and $G^{\rm np}$ is the nonpolar solvation free energy.  The related polar and nonpolar solvation  models are described in the next two sections.

\subsubsection{Polar solvation models}

We utilize either the standard Poisson model or a KSDFT based Poisson model for polar solvation free energy modeling and calculation. These two approaches are described below.

\paragraph{The Poisson model}

 The Poisson model  is given by \cite{Sharp:1990,Geng:2007a}
\begin{equation}
\label{PoissonModel}
\left\{
  \begin{array}{ll}
    -\nabla\cdot(\epsilon(\mathbf{r})\phi(\mathbf{r}))=\sum_{i=1}^{N_m}q_i\delta(\mathbf{r}-\mathbf{r}_i), & \ \forall \mathbf{r}\in\Omega \\
    \phi_{\rm s}(\mathbf{r})-\phi_{\rm m}(\mathbf{r})=0, & \ \forall \mathbf{r}\in\Gamma \\
    \epsilon_{\rm s}\nabla\phi_{\rm s}(\mathbf{r})\cdot\mathbf{n}-\epsilon_{\rm m}\nabla\phi_{\rm m}(\mathbf{r})\cdot\mathbf{n}=0, & \ \forall \mathbf{r}\in\Gamma \\
    \phi(\mathbf{r})=\frac{1}{4\pi}\sum_{i=1}^{N_m}\frac{q_i}{\epsilon_{\rm s}|\mathbf{r}-\mathbf{r}_i|}, & \ \forall \mathbf{r}\in\partial\Omega,
  \end{array}
\right.
\end{equation}
where $\epsilon(\mathbf{r})$ is the discontinuous dielectric constant and is chosen as $\epsilon_m =1 $ in the solute domain $\Omega_{\rm m}$ and at $\epsilon_s =80$ in the solvent domain  $\Omega_{\rm s}$. Here $\phi$ is the electrostatic potential, $N_m$ is the total number of partial charges in the solute, $q_i$ is the $i$th charge at the position of ${\bf r}_i$.   Additionally, $\phi_{\rm s}$ and   $\phi_{\rm m}$ are limiting values of the potential from solvent and solute domains, respectively. Here ${\bf n}$ is the  outer norm vector at point ${\bf r}$ of the interface $\Gamma$. Finally, $\Omega$ and $\partial\Omega$ are   computational domain and domain boundary, respectively.

We use the solvent excluded surface (SES) in the present work and it is generated by using our in-house  Eulerian solvent excluded surface    (ESES) software \cite{Beibeisurface:2015}. ESES  automatically embeds the SES into the Cartesian mesh, which is  convenient for the finite difference type of Poisson solvers. The matched interface and boundary based PB (MIBPB) software package \cite{BaoWang:MIBPBSolver,DuanChen:2011a,Geng:2007a} is used to solve Eq. (\ref{PoissonModel}) with appropriate charge force fields for $\{q_i\}$ and  radius sets in  SESs.

The polar solvation free energy is then calculated as
\begin{equation}
\label{PolarE1}
\Delta G^{\rm p} = \frac{1}{2}\sum_{i=1}^{N_m} q(\mathbf{r}_i)(\phi(\mathbf{r}_i) - \phi_{\rm home}(\mathbf{r}_i)),
\end{equation}
where $\phi_{\rm home}(\mathbf{r}_i)$ is the solution of the  Poisson equation with  uniform dielectric constant  $\epsilon_m$ over the whole domain $\Omega$.

\paragraph{The KSDFT based polarizable Poisson model}

Instead of using a given  partial charge force field,  the KSDFT based polarizable Poisson model utilizes a DFT approach to compute the charge distribution in the solvent environment.  As such, this approach partially takes into account for solvent polarization and solute response in an iterative procedure. The combination of the Poisson equation with SES and KSDFT  has been reported in the literature \cite{MLWang:2006, MLWang:2007}. Recently, we have proposed a differential geometry based polarizable PB model that further allows solvent-solute interface to be optimized in the Poisson-Kohn-Sham (PKS) formalism \cite{ZhanChen:2011a}. In the present work, we adopt SES in our PKS software for the sake of an accurate estimation of atomic surface areas via our ESES software package  \cite{Beibeisurface:2015}.   Consequently, our MIBPB software is also employed for solving the Poisson equation with sharp interfaces \cite{BaoWang:MIBPBSolver,DuanChen:2011a,Geng:2007a}.

In this approach,  the Poisson model is  \cite{ZhanChen:2011a}
\begin{equation}
\label{PPoissonModel}
\left\{
  \begin{array}{ll}
    -\nabla\cdot(\epsilon(\mathbf{r})\phi(\mathbf{r}))= {\rho}_{\rm total}(\mathbf{r}), & \ \forall \mathbf{r}\in\Omega \\
    \phi_{\rm s}(\mathbf{r})-\phi_{\rm m}(\mathbf{r})=0, & \ \forall \mathbf{r}\in\Gamma \\
    \epsilon_{\rm s}\nabla\phi_{\rm s}(\mathbf{r})\cdot\mathbf{n}-\epsilon_{\rm m}\nabla\phi_{\rm m}(\mathbf{r})\cdot\mathbf{n}=0, & \ \forall \mathbf{r}\in\Gamma \\
    \phi(\mathbf{r})=
		\frac{1}{4\pi}\int_\Omega \frac{\rho_{\rm total}(\tilde{\mathbf{r}})}{\epsilon_{\rm s}|\mathbf{r}-\tilde{\mathbf{r}}|}d\tilde{\mathbf{r}}, & \ \forall \mathbf{r}\in\partial\Omega,
  \end{array}
\right.
\end{equation}
where  ${\rho}_{\rm total}(\mathbf{r})=q n(\mathbf{r})-qn_n(\mathbf{r})$ is the charge distribution in the solute domain, with $q$ being the unit charge of an electron, $n(\mathbf{r})$ being the electron density, and $n_n(\mathbf{r})=\sum_IZ_I\delta(\mathbf{r}-\mathbf{R}_I)$ being the nucleus density, in which $Z_I$ and $\mathbf{R}_I$ are the atomic number and position vector of nucleus $I$, respectively.

The Poisson equation  (\ref{PPoissonModel}) is coupled to  the following generalized  KS  equation \cite{ZhanChen:2011a},
\begin{equation}
\label{PGKohnShamS}
\left(-\frac{\hbar^2}{2m}\nabla^2+U_{\rm eff}\right)\psi_i=E_i\psi_i,
\end{equation}
where $m$ is electron mass and  $\hbar^2=h/2\pi$ with $h$ being the Planck constant. Here $\psi_i $ and $E_i$ are the $i$th KS orbital  and $i$th eigenvalue, respectively, and  $U_{\rm eff}\doteq q\phi_{\rm RF}+U_{\rm eff}^0$ is the effective potential of the generalized Kohn Sham equation with $\phi_{\rm RF}=\phi-\phi_0$ being the reaction field potential of the solute in the solvent environment, in which $\phi$ and $\phi_0$ are the electrostatic potential generated by the solute in the solvent and solute environments, respectively. Finally, $U_{\rm eff}^0$ is the effective Kohn Sham potential in the vacuum environment. Note that the solution of Eq. (\ref{PGKohnShamS}) gives rise to the electron density $n(\mathbf{r})=\sum_i|\psi_i|^2$, which, in turn, updates the charge distribution for  Eq. (\ref{PPoissonModel}). A  self-consistent iterative process has been developed to solve Eqs. (\ref{PPoissonModel}) and (\ref{PGKohnShamS}) to their convergence \cite{ZhanChen:2011a}  starting from solving the KSDFT in vacuum as an initial guess to the coupled system.

%Detail of SIESTA implementation and communication between two solvers
For KSDFT calculations, we  employ the SIESTA solver \cite{SIESTA:2002}, which uses pseudopotential to eliminate the complicated effects of core electrons. In all calculations, the default double-$\zeta$ plus single polarization (DZP) basis is used. The MeshCutOff is set as 125 Rydberg and local density approximation (LDA) is used to approximate the exchange correlation potential. The solution method is set to be ``diagon''.

In this  formalism, the polar solvation free energy is computed by
\begin{equation}
\label{PolarEnergyPPoisson}
\Delta G^{\rm p}=\int_{\Omega^{\rm m}}q(\mathbf{r})\phi_{\rm RF}(\mathbf{r})d\mathbf{r}.
\end{equation}

\subsubsection{Nonpolar solvation models}

In our earlier work, the nonpolar solvation free energy was modeled by \cite{Wei:2009, ZhanChen:2011a, ZhanChen:2010a, ZhanChen:2010b}
\begin{equation}
\label{NonPolarEnergyOld}
 G^{\rm np}=\gamma A + pV +  +\rho_0 \int_{\Omega_s} U^{\text{vdW}} \mathrm{d}\mathbf{r},
\end{equation}
where  $\gamma$ and  $A$   are, respectively, surface tension and area of the solute molecule. Additionally, $p$ and $V$  are, respectively  surface enclosed volume  and hydrodynamic pressure difference. Finally, $\rho_0$ is the solvent bulk density, and $U^\text{vdW}$ is the vdW interaction potential, i.e., the Lennard-Jones (LJ) potential. The  integration is    over solvent domain $\Omega_s$. This nonpolar solvation free energy model was shown to offer  excellent predictions of experimental data for various nonpolar molecules \cite{ZhanChen:2012}. In our recent work, we have demonstrated superb results for a large number of polar and nonpolar molecules  when the nonpolar model in Eq. (\ref{NonPolarEnergyOld}) is combined with a PB model for the polar solvation component \cite{BaoWang:2015a}.

The LJ potential offers a physical model for dealing with vdW interactions near the solvent-solute interface. However, a drawback of this nonpolar term is that the probe radius in the LJ  potential is nonlinear and cannot be optimized together with other nonpolar parameters. Mathematically, for a small probe radius, the vdW term for a given atom is proportional to the atomic surface area. Therefore, atomic surface area approach used by Kollman and co-workers  \cite{Kollman:2001Solvation} should have a similar modeling effect as that of the LJ potential. Based on this observation, we propose the following nonpolar solvation energy model
\begin{equation} \label{NonPolarEnergy}
\Delta G^{\rm np}=\sum_{\alpha =1}^N  \gamma^\alpha {\rm Area}^\alpha+p{\rm Vol}=\sum_{\alpha =1}^N \sum_{i\in \alpha }\gamma^\alpha {\rm Area}_i^\alpha+p{\rm Vol},
\end{equation}
where $N$ is the total number of atomic types in a given solute molecule. Here, $\gamma^\alpha$ and ${\rm Area}^\alpha$ are the surface tension and surface area of the $\alpha$th type of atoms, respectively,  and ${\rm Area}_i^\alpha$ is the surface area of the $i$th atom in the $\alpha$ type of atoms. In this nonpolar solvation free energy model, the parameters $\gamma_\alpha$ and $p$ need to be learned from a training set. Both the LJ and the atomic-surface-area based nonpolar solvation models are validated in the present work.

\subsection{Functional group  classification}\label{NonpolarEnergyModel1}

%Nonpolar solvation models are described in this section.
%In the present work, the nonpolar solvation free energy is modeled by atomic surface areas and surface enclosed volume. For a given functional group molecules, we learn a set of parameters from the training set and further using these parameters to predict the solvation free energy of the same functional group molecules. For the poly-functional groups compounds, the functional groups are scored by using the similar poly-functional group molecules. The sets of parameters and scoring weights are combined to predict the solvation free energy. Detailed methodology will be described in the following parts of this section.

\paragraph{Functional group modeling}

In our nonpolar solvation model, we assume that each functional group of molecules admits the same set of optimal parameters. Similar approach has been   successfully  used in  the literature \cite{Sherman:2010}, including ours \cite{ZhanChen:2012,BaoWang:2015a}. In this work, we further incorporate machine learning type of methods for the nonpolar solvation free energy prediction.  To this end, the whole data set, which contains 668 molecules, is partitioned into  training sets and  testing sets. In the leave one out test case, each step we only leave one molecule out as the test set, all the other molecules are regarded as the training set. In a specific leave-SAMPLx-out study, all the SAMPLx molecules are left as the testing set, while the remaining molecules are treated as the training set. Furthermore, the training set in this scenario also divided into two parts: i) the mono-functional group molecules, in which molecules of a specific functional group is used to train a set of parameters for all similar molecules; and ii) the poly-functional groups molecules in which scoring weights are used to weight the contributions of parameters of various involved functional groups.
%molecules with  involved classes of functional groups  are used to obtain scoring weights between different functional groups.
%In the blind prediction step, the parameters learned from the training set will be employed for the prediction of the molecules in the testing set.

%List functional groups selected
\begin{table}[!ht]
\centering
\caption{Functional groups and corresponding number of molecules used in the classification.}
\begin{tabular}{llllllll}
\hline
{\rm Group} &{\rm Number}  &{\rm Group}  &{\rm Number} &{\rm Group} &{\rm Number}  &{\rm Group} &{\rm Number} \\
\hline
alkynyl  &8	&alkenyl	&38	&aldehyde group	&11  &nitrile group  &5\\
carboxyl &7	&ester group	&34	&ketone	&23  &amino  &35\\
nitro    &9 &alcoholic hydroxy    &33    &phenolic hydroxyl   &16   &ether   &22  \\
alkane  & 38 &aromatics    &33   &nitrogen heterocyclic  &19  &chlorinated hydrocarbon  &53 \\
Nitrate   &5   &amid   &7  &thiol  &4  &thioether  &5 \\
\hline
\end{tabular}
\label{ClassesOfMolecules}
\end{table}

Table \ref{ClassesOfMolecules} lists twenty functional groups used in the training set. We denote  $\{T_1, T_2, \cdots, T_n\}$ a given set of $n$ molecules with a specific functional group  from the above 20 groups. For a molecule indexed  $j$, the solvation free energy calculated from  the polar solvation free energy model,  atomic surface areas and molecular volume are listed as:
\begin{equation}
\label{MolFeatures}
\left\{\Delta G_j^{\rm p}, {\rm Area}_j^1, {\rm Area}_j^2, \cdots, {\rm Area}_j^N, {\rm Vol}\right\},
\end{equation}
where $j=1, 2, \cdots, n$.

In our approach, the solvation free energy is modeled as the sum of the polar and nonpolar parts, thus for the $j$th molecule, the corresponding modeled solvation free energy can be expressed as:
\begin{equation}
\Delta G_j=\Delta G_j^{\rm p}+\sum_{\alpha=1}^N \gamma^\alpha {\rm Area}^\alpha+p{\rm Vol}\doteq G_j(\mathbf{P}),
\end{equation}
where $\mathbf{P}\doteq \{\gamma^1, \gamma^2, \cdots, \gamma^N, p\}$ is the set of parameters to be optimized.

For a given set of molecules with the same mono-functional group, we denoted $\Delta \mathbf{G}(\mathbf{P})\doteq\left\{ G_1(\mathbf{P}), G_2(\mathbf{P}), \cdots, G_n(\mathbf{P}) \right\}$, and the associated experimental solvation free energy is denoted as $\Delta G^{\rm Exp}\doteq\left\{\Delta G_1^{\rm Exp}, \Delta G_2^{\rm Exp}, \cdots, \Delta G_n^{\rm Exp} \right\}$.  Then the optimal parameter set can be obtained by solving the following Tikhonov regularized least square problem (also known as ridge regression) which has a closed form solution:
\begin{equation}
\label{TRLS}
{\rm argmin}_{\mathbf{P}}\left\{||\Delta G(\mathbf{P})-\Delta G^{\rm Exp}||_2+\lambda ||\mathbf{P}||_2\right\},
\end{equation}
where $||*||_2$ is the $L_2$ norm of the quantity $*$. Here $\lambda$ is the regularization parameter chosen to be a large number, such as 100, in the present work to ensure the dominance of the first term and avoid over-fitting through controlling the magnitude of $||\mathbf{P}||_2$..

\paragraph{Functional group scoring}
Most molecules in the SAMPL blind test sets involve poly-functional groups. In this case, we further employ the poly-functional group molecules in the training set for training the optimal relative scoring weights between different functional groups. According to the relative scoring weights, the scoring weights between all the functional groups can be obtained through a simple normalization procedure. We denote $\left\{\tilde{T}_1, \tilde{T}_2, \cdots, \tilde{T}_{n'}\right\}$ a given set of poly-functional group molecules that has the same functional groups in the training set. The associated optimized parameter sets are $\mathbf{P}_1, \mathbf{P}_2, \cdots, \mathbf{P}_m$, where $m$ is the number of functional groups in this set, each $\mathbf{P}_i$, $i=1, 2, \cdots, m$ is learned through solving the regularized optimization problem given by Eq. (\ref{TRLS}). For the $j$th molecule in this poly-functional group  set, we model its solvation free energy as:
\begin{equation}
\label{SolEngPoly}
\Delta \bar{G}_j(\boldsymbol{\omega})=\sum_{i=1}^m \omega_i \Delta G_j(\mathbf{P}_i),
\end{equation}
where $||\boldsymbol{\omega}||_1\doteq \sum_{i=1}^m \omega_i=1$, with $\omega_i$ being the scoring weight   of $i$th functional group.

The relative scoring weights for the $m$ functional groups associated to this set of poly-functional groups molecules can be learned via solving the following constraint optimization problem,
\begin{equation}
\label{ScoringWeightsOptimal}
{\rm argmin}_{\boldsymbol{\omega}} ||\Delta \tilde{G}(\boldsymbol{\omega})-\Delta\tilde{G}^{\rm Exp}||_2,
\end{equation}
subject to
\begin{equation}
\label{ScoringWeightsOptimalConstraint}
||\boldsymbol{\omega}||_1=1,
\end{equation}
and
\begin{equation}
\label{ScoringWeightsOptimalConstraint2}
\omega_i\geq 0, \ \forall i=1, 2, \cdots, m,
\end{equation}
where $\Delta \tilde{G}(\boldsymbol{\omega})$ and $\Delta\tilde{G}^{\rm Exp}$ represent, respectively, the predicted and experimental solvation
free energies for this group of poly-functional group molecules. Since both the optimization object given by Eq.(\ref{ScoringWeightsOptimal}) and the constraint conditions in Eqs. (\ref{ScoringWeightsOptimalConstraint}-\ref{ScoringWeightsOptimalConstraint2}) are convex with respect to the scoring weights $\boldsymbol{\omega}$. The above constrained optimization can be easily solved via a convex optimization solver in the CVX software package \cite{cvx, gb08}.

%List the transition by an example
\begin{figure}[!ht]
\small
\centering
\includegraphics[width=10cm,height=12cm]{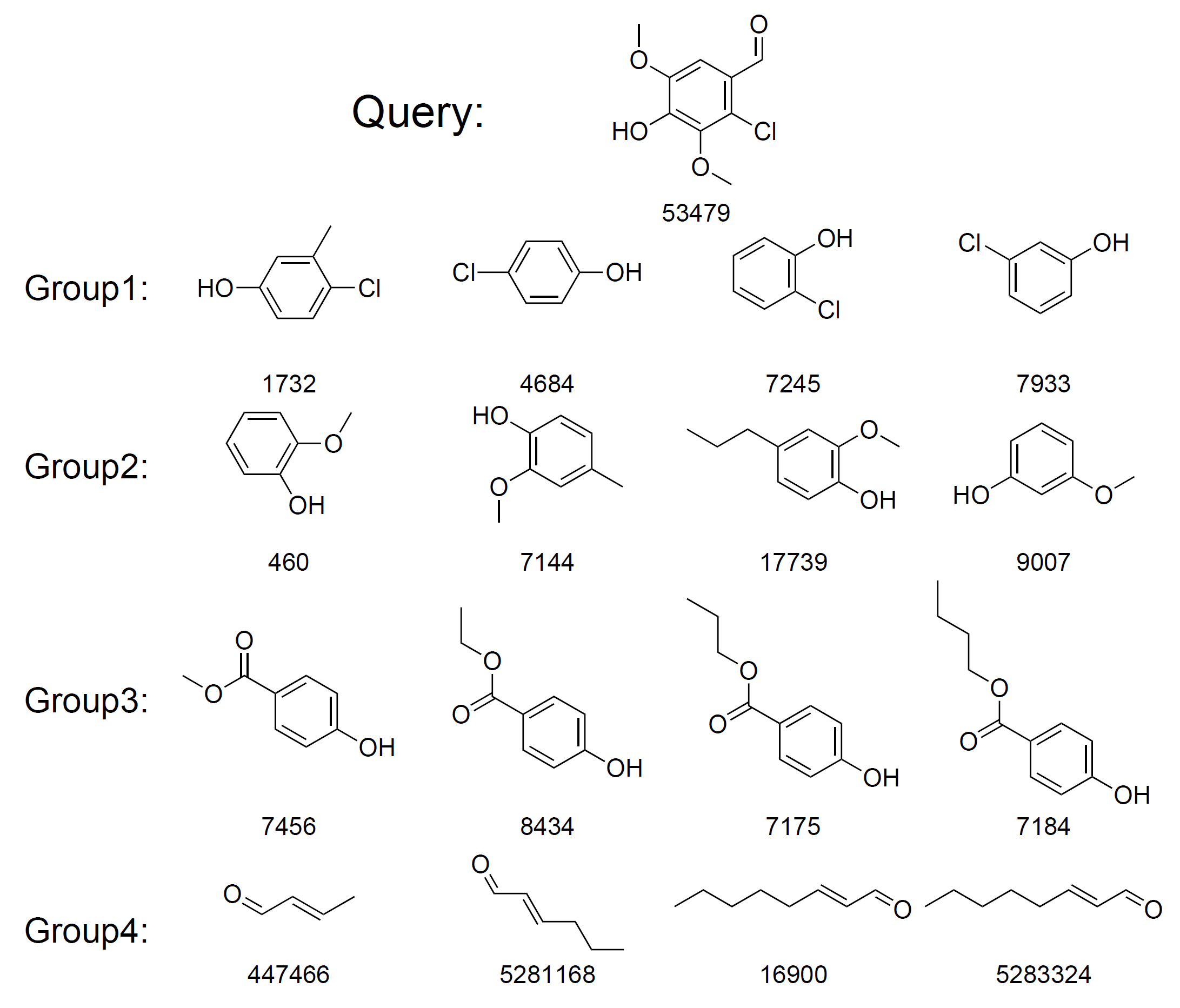}
\caption{An illustration of the functional group scoring method for the prediction of the solvation free energy for molecule 2-Chlorosyringaldehyde (Pubchem ID: 53479), which contains four functional groups: aldehyde group, phenolic hydroxyl, ether and chlorinated hydrocarbon. We first compute relative weights between phenolic hydroxyl and chlorinated hydrocarbon; phenolic hydroxyl and ether; phenolic hydroxyl and ester group; and ester group and aldehyde group. Then relative weights are combined to generate the full set of  weights $\omega_1, \omega_2, \omega_3$ and $\omega_4$  for solvation free energy prediction.}
\label{FunctionalGroupScoring}
\end{figure}

In the  rest of this section, we  provide an example to illustrate the procedure of functional group based approach for solvation free energy prediction. As shown in Fig. \ref{FunctionalGroupScoring}, the target molecule 2-Chlorosyringaldehyde (Pubchem ID: 53479) contains four different functional groups.  We need to find   scoring weights for these functional groups. Note that in the above molecular searching scheme, pairwisely, the relative weights for each two functional groups can be determined by solving the constrained optimization problem in Eqs. (\ref{ScoringWeightsOptimal})-(\ref{ScoringWeightsOptimalConstraint}) for molecules in the two corresponding functional groups. According to the pairwise relative weights, the functional group scoring in the target molecule can be achieved.

\subsection{Nearest neighbor search  }\label{NonpolarEnergyModel2}

\begin{figure}[!h]
\small
\centering
\includegraphics[width=4cm,height=2cm]{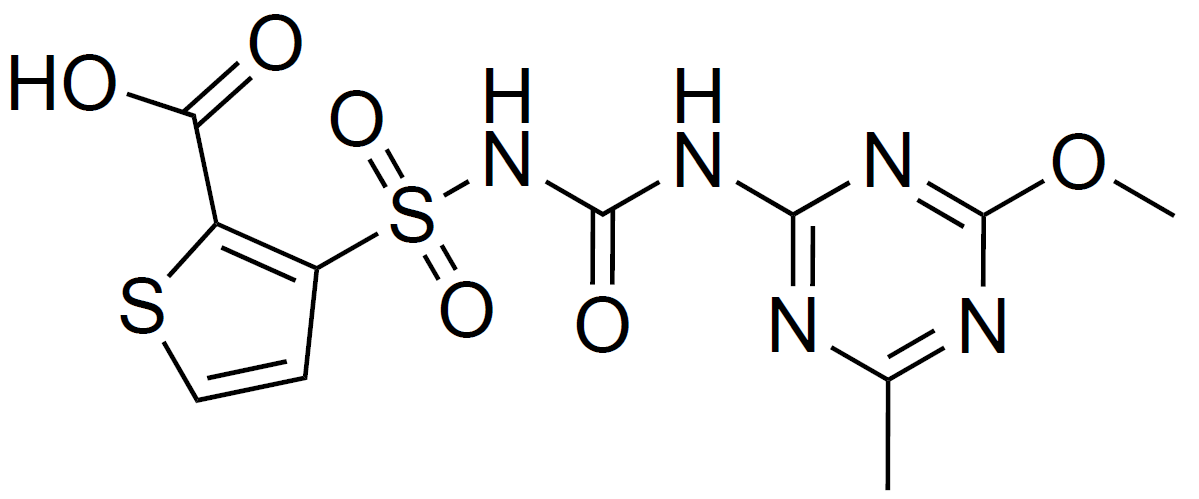}
\caption{Thifensulfuron (Pubchem ID: 91729).   }
\label{FunctionalGroupFailed}
\end{figure}

The above functional group based model is tested to be able to provide very accurate solvation free energy prediction provided that for each test molecule, parameters for all of the involved mono-functional groups can be determined, and a set of molecules with the same ploy-functional groups can be found in the training set as well. However, for some complex molecules that contain functional groups beyond those listed in Table \ref{ClassesOfMolecules}, this method fails. For instance, considering a  SAMPL1   test molecule,  thifensulfuron, as shown in Fig. \ref{FunctionalGroupFailed}, it has a very complex structure  and contains functional groups cannot be found in Table \ref{ClassesOfMolecules}. We therefore propose the following ranking algorithm.

\paragraph{Atomic feature based molecular ranking}

To deal with general and complex molecules,   we propose another approach to rank molecules, and develop a nearest neighbor approach for nonpolar solvation free energy prediction. Our molecular ranking  is based on  atomic features. Our basic assumption is that if two solute molecules have similar atomic features, their chemical properties are also similar. Therefore, these molecules should share the same parameter set for the nonpolar solvation free energy model. This approach is further described  below.

 Our method is based on the ansatz that molecules chemical properties are mainly determined by atomic features, including structural features and atomic electrostatic features.  Among them, atomic structural features include atomic type, atom  hybridization state and bonding information. Atomic electrostatic features include atomic charge, atomic dipole,  atomic quadrupole and  atomic electrostatic solvation free energy.
Atomic features are selected based the criteria that a given atomic feature is retained if it can effectively discriminate  the previously mentioned mono-functional groups listed in Table \ref{ClassesOfMolecules}.   Table \ref{AtomicFeatures} lists all the atomic features that are used for  molecule ranking.

\begin{table}[!ht]
\centering
\caption{Atomic features used for ranking molecules.}
\begin{tabular}{ll}
\hline
{\rm Feature name} & {\rm Feature name}\\
\hline
Number of atoms  & Number of heavy atoms\\
Number of hydrogen atoms & Number of single bonds	\\
Number of double bonds  & Number of triple bonds\\
Number of aromatic bonds	& Number of each type of atoms\\
Number of  $sp^1$ carbon atoms & Number of $sp^2$ carbon atoms\\
Number of  $sp^3$ carbon atoms & Number of $sp^1$ nitrogen atoms\\
Number of  $sp^2$ nitrogen atoms & Number of $sp^3$ nitrogen atoms\\
Number of  $sp^2$ oxygen atoms & Number of $sp^3$ oxygen atoms\\
Number of  $sp^2$ sulfur atoms & Number of $sp^3$ sulfur atoms\\
Maximum of atomic reaction field energy	& Minimum of atomic reaction field energy\\
Maximum reaction field energy of each type of atoms & Minimum reaction field energy of each type of atoms\\
Maximum of atomic reaction field energy & Average reaction field energy of each type of atoms\\
Total absolute charge & Total charge of each type of atoms \\
Total absolute charge of each type of atoms & Maximum charge of each type of atoms \\
Minimum charge of each type of atoms &The variation of each type of atomic charges\\
Maximum of atomic dipole & Maximum of each atomic dipole \\
Minimum of each atomic dipole & Variation of atomic dipole \\
Maximum quadrupole & Maximum quadrupole of each type of atoms\\
Variation of atomic quadrupole & Variation of each type of atom's quadrupole \\
\hline
\end{tabular}
\label{AtomicFeatures}
\end{table}
We also construct  atomic features by using the statistics variables, i.e., average, maximum, minimum, and variation of quantities in Table \ref{AtomicFeatures}.
%Finally, we rule out all the features that are correlated  with  previously selected atomic features.
Finally, we rule out  redundant features, where redundancy is due to the fact that some atomic features are 100 $\%$ correlated with each other.  In this case, only one of these highly correlated features is retained.

Atomic features are calculated by  following methods.
\begin{itemize}
\item Molecular structural information is parsed by the Open Babel software \cite{pybel}.
\item Atomic charge, atomic dipole and atomic quadrupole are obtained via the distributed multipole analysis (DMA) method \cite{DMA:1981}, in which the charge density is originally computed by the density function theory with B3LYP and 6-31G basis   in the Gaussian quantum chemistry software \cite{g09, B3, LYP}.
\item Atomic electrostatic solvation energy is calculated by our in-house MIBPB software \cite{BaoWang:MIBPBSolver,DuanChen:2011a,Geng:2007a}.
\end{itemize}

With the above selected atomic features, intuitively, we measure the similarity of molecules by the Pearson correlation coefficient of  atomic feature vectors. Specifically, we denote the features of two molecules as vector $X\doteq \{x_1, x_2, \cdots, x_k\}$ and vector $Y\doteq \{y_1, y_2, \cdots, y_k\}$, then their similarity is measured by
$$
C_{XY}=\frac{|\sum_{i=1}^k(x_i-\bar{x})(y_i-\bar{y})|}{\sqrt{\sum_{i=1}^k (x_i-\bar{x})^2}\sqrt{\sum_{i=1}^k (y_i-\bar{y})^2}},
$$
where $\bar{x}\doteq \frac{1}{k}\sum_{i=1}^k x_i$ and $\bar{y}\doteq \frac{1}{k}\sum_{i=1}^k y_i$, and $k$ is the dimension of the feature space. The higher the correlation between two atomic feature vectors indicates the more similarity between molecules.

\paragraph{Nearest neighbor solvation free energy prediction}

By using the above  nearest neighbor based nearest molecule searching procedure, for a given molecule, we obtain the similarity ranking of the remaining molecules with this specified one, as well as correlations between molecules. In the nearest neighbor prediction of the nonpolar solvation free energy, we learn the parameters in the nonpolar solvation free energy expression, Eq. (\ref{NonPolarEnergy}), by a given number of nearest molecules found in the training set. The number of the nearest molecules used is based on the following principles
\begin{itemize}
\item All the molecules whose correlation with a given molecule is greater than 0.99 are used for the parameter learning for this given molecule;
\item If the above criterion yields  less than 5 molecules, we use 5 nearest molecules for the parameter learning. Here the 5-molecule nearest neighbor method  is found to provide the best leave-one-out prediction.
\end{itemize}

\subsection{Unified protocol  for blind solvation free energy prediction} \label{SolvationModel}

Models for the total solvation free energy are given in Eq. (\ref{TotalModels}). We utilize a unified protocol for the prediction of solvation free energy. First, for the polar solvation free energy  $\Delta G^{\rm p}$, we select either the Poisson model with a given point charge force field or the polarizable Poisson model. Second, for the calculation of the nonpolar solvation free energy we note that in general, the functional group scoring approach can deliver better blind predictions than the nearest neighbor approach.  Therefore, in our nonpolar energy prediction step, the functional group scoring method is used whenever it works. Otherwise, we utilize the nearest neighbor approach.

\section{Results and discussions}\label{NumericalResults}

\subsection{Date processing and model validation}

\paragraph{Data sets and force fields}

We consider a total of 668 molecules, the structures of these molecules are downloaded from the \href{https://pubchem.ncbi.nlm.nih.gov/}{Pubchem} project. The experimental solvation free energies of these molecules are collected from the literature. This dataset contains   molecules from the Sampl blind solvation prediction projects, ranging from SAMPL0 to SAMPL4 \cite{nicholls2008predicting, SAMPL1:2009, SAMPL:2010, SAMPL3:2012, Guthrie:2014solvation}; and the remaining molecules in our data set are collected from the literature \cite{Cabani:1981, Kollman:2001Solvation, Truhlar:1999Solvationext}. Coincidentally, there is a considerable overlap of our database with Mobley's solvation database, which is available from \url{http://mobleylab.org/resources.html}. More precisely, 589 molecules in our database are already covered by Mobley's. The detail information of our dataset is provided in the Supporting material \cite{Supporting}.

For both the standard  Poisson model and the KSDFT based polarizable Poisson model, we consider four types of atomic radii, namely, Amber 6, Amber bondi, Amber mbondi2 \cite{AMBER15} and ZAP9 \cite{Nicholls:2008solvation} parametrizations. Additionally, for the standard Poisson model, three sets of charge assignments, namely, OpenEye-AM1-BCC v1 parameters \cite{Jakalian:2000}, Gasteiger \cite{Gasteiger:1980}, and Mulliken \cite{AMBER15}, are tested. Our MIBPB solver \cite{BaoWang:MIBPBSolver,DuanChen:2011a,Geng:2007a} is utilized to solve the Poisson interface problem to obtain electrostatic solvation energies. The probe radius is set to 1.4\AA ~ in all PB calculations.

For the KSDFT based polarizable Poisson model, the Poisson interface problem is coupled with the generalized KSDFT in a self-consistent approach. The charge density used by Poisson interface problem is calculated in {\it Ab Initio} approach by the generalized KSDFT, here generalization due to the additional solute-solvent reaction field energy is included in the effective KS potential. The Poisson interface problem is employed for calculating the reaction field energy. This approach is modified from our earlier differential geometry based polarizable PB model \cite{ZhanChen:2011a}. We   iteratively couple the SIESTA \cite{SIESTA:2002} and the MIBPB solver \cite{BaoWang:MIBPBSolver,DuanChen:2011a,Geng:2007a}, in which a sharp solvent-solute interface is employed. The SIESTA software with additional reaction field energy is used for charge density calculation, and our in-house software is used for reaction field energy calculation. A uniform mesh size of 0.25 \AA~ is used for solving the Poisson equation in both the standard Poisson and Polarizable Poisson model. In the polarizable Poisson model, the computed change densities are  mapped to the uniform mesh with the conservation of the total charge.

%Data used in the present computation  including molecular names, PubChem IDs, coordinates (download from the PubChem database), atomic radii, charges and experimental solvation free energies are available in the Supporting material \cite{Supporting}.

%Table \ref{OmittedMols} lists four molecules that are omitted in SAMPL blind test sets.

\begin{table}
\centering
\caption{Molecules in   SAMPLx sets involving   bromine and/or iodine atoms.}
\begin{tabular}{ c|c }
%\hline
%\multicolumn{6}{ c }{The RMS error of the solvation free energy prediction with different method} \\
\hline
Test set                & Molecule \\ \hline
\multirow{1}{*}{SAMPL0} & benzyl bromide \\
\hline
\multirow{1}{*}{SAMPL1} & bromacil\\
\hline
\multirow{2}{*}{SAMPL2} & 5-bromouracil\\
                        & 5-iodouracil\\
\hline
\multirow{1}{*}{SAMPL3} & None\\
\hline
\multirow{1}{*}{SAMPL4} & None \\
\hline
\end{tabular}
\label{OmittedMols}
\end{table}

%We consider a total of 668 molecules. Most of them (about 628??) are downloaded from Mobley's database \cite{Mobley:2014} available from  Mobley's homepage \url{http://mobleylab.org/resources.html}. The rest of molecules are extracted from the literature \cite{Cabani:1981}.
%In fact, Mobley's database \cite{Mobley:2014} has 643 molecules.  (More details with respect to Mobley's data is needed!!! Exactly, what molecules are excluded??)

For the molecules containing iodine atoms, the current level of DFT method used in this work, including the Gaussian software, cannot handle this element appropriately. Therefore, for a uniform comparison, we ignore molecules containing  iodine atoms.  There is a similar situation for  bromine atoms --- there is no appropriate pseudo-potential for this atom in the SIESTA software. Therefore, molecules involving bromine atoms are excluded in our KSDFT based polarizable PB calculations. Table \ref{OmittedMols} lists four molecules in SAMPLx that are not considered in some of our predictions.

\paragraph{Atomic surface area and molecular volume calculation}

In our nonpolar solvation free energy model, both atomic areas and surface enclosed volume are required. In a recently developed Eulerian solvent excluded surface (ESES) package, a second-order accurate numerical scheme for surface area calculations and a third-order accurate numerical scheme for volume estimations have been developed \cite{Beibeisurface:2015}. A weighted Voronoi diagram algorithm is implemented to partition a molecular surface area into atomic surface areas. These schemes have been intensively validated by a large number of  test examples. In this work, both atomic areas and molecular volume are calculated directly by using our ESES software package \cite{Beibeisurface:2015}.

\paragraph{Validation of atomic surface area based nonpolar model}

In this work, instead of using the classical nonpolar model that includes the van der Waals interaction between the solvent and solute nonpolar interaction \cite{BaoWang:2015a} as shwon in Eq. (\ref{NonPolarEnergyOld}), we utilize the atomic surface area model given in Eq. (\ref{NonPolarEnergy}) for the corresponding effects. In this part, we consider  a set of numerical test to verify the validity of this treatment.
Our numerical results indicate that the atomic surface area model  provides results as good as those obtained by the van der Waals based nonpolar model.

\begin{table}
\centering
\caption{The RMS errors of the solvation free energy prediction with atomic surface area and van der Waals interaction models of nonpolar solvation free energy for SAMPL0 test set. The molecule in the SAMPL0 set that contains Br atom is excluded from this comparison. All results are in unit kcal/mol}
\begin{tabular}{ c|l|cccc }
%\hline
%\multicolumn{6}{ c }{The RMS error of the solvation free energy prediction with different method} \\
\hline
Nonpolar model & Radius &BCC & Mulliken & Gasteiger &SIESTA \\ \hline
\multirow{4}{*}{Atomic surface area} &Amber 6        &1.30   &1.27   &1.23  &0.99 \\
                                     &Amber Bondi    &1.40   &1.30   &1.30  &0.93 \\
                                     &Amber Mbondi2  &1.41   &1.34   &1.33  &1.10 \\
                                     &ZAP9           &1.33   &1.39   &1.37  &1.08 \\
\hline
\multirow{4}{*}{van der Waals} &Amber 6        &1.42   &1.31  &1.34 &0.93 \\
                               &Amber Bondi    &1.44   &1.30  &1.30 &1.05 \\
                               &Amber Mbondi2  &1.45   &1.38  &1.30 &1.18 \\
                               &ZAP9           &1.44   &1.39  &1.32 &1.32 \\
\hline
\end{tabular}
\label{S0Comparison-VDW-AtomicArea}
\end{table}

The van der Waals interaction  Eq. (\ref{NonPolarEnergyOld}) is modeled  as a semi-continuous and semi-atomic potential. More specifically, we consider the 6-12 LJ potential  for modeling the van der Waals interaction between the continuum solvent and atomistic solute molecules \cite{BaoWang:2015a}.  The probe radius of of the solvent in the LJ potential is set to be 1.4 \AA, which is the same as that used in all other calculations.
Table \ref{S0Comparison-VDW-AtomicArea} lists the root mean square (RMS)   errors with four types of atomic radii and four charge parametrization methods for the prediction of SAMPL0's solvation free energies. From these results, it is seen that  for a given force field parametrization, the atomic surface area approach provides slightly more accurate nonpolar solvation free energy prediction. As shown in our earlier work \cite{BaoWang:2015a}, the performance of the semi-continuous and semi-atomic LJ potential is sensitive to the choice  of the probe radius, due to its nonlinear dependence on the radius, whereas the SES based atomic surface area approach is less sensitive to the probe radius. We conclude  that in general the atomic surface area based nonpolar model is at least as good as the LJ potential based one  for modeling the nonpolar solvation free energy.   Therefore,  the atomic surface area based nonpolar model is employed in the rest of this paper.

\subsection{Solvation predictions}

%%%%%%%%%%%%%%%%%%%%%%%%%%%%%%%%%%%%%%%%%%%%%%%%%%%%%%%%%%%%%%%%%%%%%%%%%%%%%%%%%%%
%Leave one out
%%%%%%%%%%%%%%%%%%%%%%%%%%%%%%%%%%%%%%%%%%%%%%%%%%%%%%%%%%%%%%%%%%%%%%%%%%%%%%%%%%%
\begin{figure}[!h]
\small
\centering
\includegraphics[width=12cm,height=8cm]{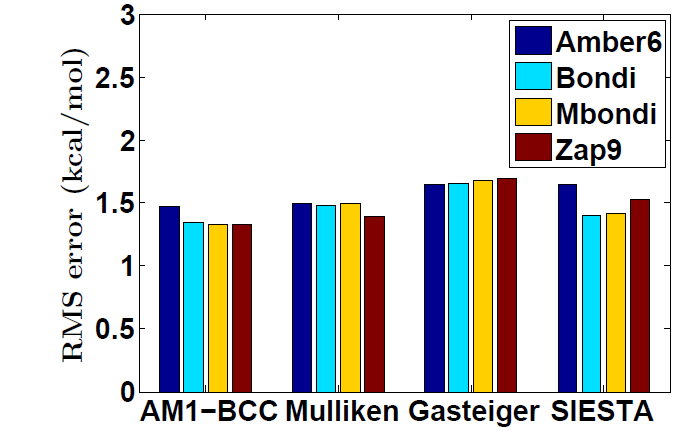}
\caption{The leave-one-out error of the whole training set of 668 molecules with 16 different   charge  and radius  parametrizations. The nearest neighbor approach is employed for solvation free energy prediction.}
\label{LeaveOneOut}
\end{figure}

\begin{figure}
\begin{center}
\begin{tabular}{cc}
\includegraphics[width=0.5\textwidth]{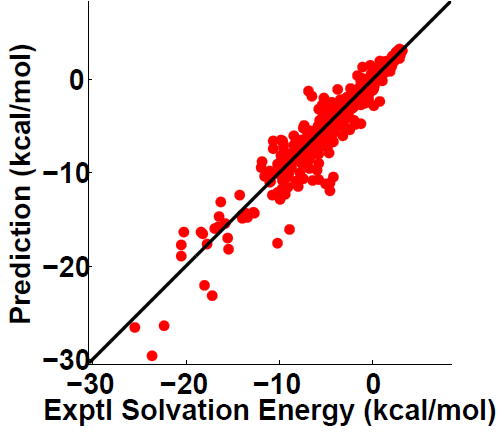}
\includegraphics[width=0.5\textwidth]{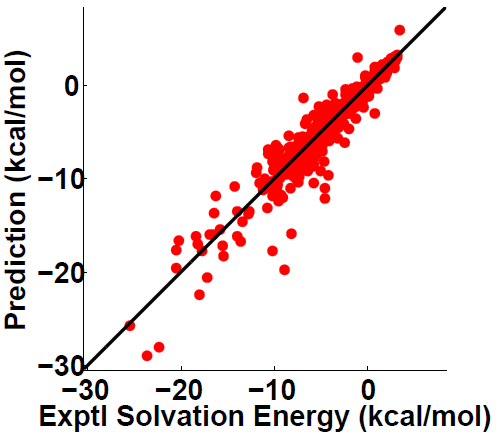}
\end{tabular}
\end{center}
\caption{Correlations of  optimal leave-one-out test results for 668 solute molecules.
Left:  Results parametrized with BCC charges and MBondi2 radii.   Pearson correlation is 0.956 and  $R^2$ is 0.913.
Right:  Results parametrized with BCC charges and ZAP9.   Pearson correlation is 0.955 and
$R^2$ is 0.911.
In both cases,  RMS errors are 1.33 kcal/mol.}
\label{LeaveOneOut-Optimal}
\end{figure}

\begin{table}
\centering
\caption{The RMS errors of the leave-one-out test of the solvation free energy prediction with different methods, all with unit kcal/mol}
\begin{tabular}{ |l|cccc| }
\hline
     Radius    &BCC   &Mulliken &Gasteiger &SIESTA \\ \hline
Amber 6        &1.47  &1.49     &1.65 &1.65 \\
Amber Bondi    &1.34  &1.48     &1.66 &1.40 \\
Amber Mbondi2  &1.33  &1.49     &1.68 &1.42 \\
ZAP9           &1.33  &1.39     &1.70 &1.53 \\
\hline
\end{tabular}
\label{LeaveOneOutTable}
\end{table}

\paragraph{Leave-one-out prediction of 668 molecules}
We first examine the proposed nearest neighbor approach by the leave-one-out test of 668 molecules. In this examination,  we select one molecule at a time and use all other molecules as the training set to predict the selected one's solvation free energy. This process is applied systematically to all the molecules in the whole dataset of 668 molecules. Four different atomic radius sets are considered, together with four different charge force fields.  Our results are illustrated in Fig. \ref{LeaveOneOut}, the associated values are listed in Table \ref{LeaveOneOutTable}.  In general, all radius sets and charge force fields perform similarly well. The maximum RMS error is below 1.7 kcal/mol for all methods over all molecules. More specifically, Bondi and Mbondi radii offer the best overall results. For charge force field AM1-BBC charges appear to provide the best predictions and their RMS errors are less than 1.5kcal/mol for all the four radius sets. The optimal result obtained with AM1-BBC charges and ZAP9 radii has an RMS error of 1.33kcal/mol. Figure \ref{LeaveOneOut-Optimal} plots the optimal results on the leave one out test.

For a set of 643 molecules, which largely overlaps with present dataset, Mobley  and  Guthrie reported an RMS error of 1.51 kcal/mol \cite{Mobley:2014}. Our results indicate that the present nearest neighbor approach can achieve highly accurate predictions for the solvation free energies for all the atomic radius and charge force fields.
%Test Out leave one out results on the same set later, (To be done, 22-04-2016)

%%%%%%%%%%%%%%%%%%%%%%%%%%%%%%%%%%%%%%%%%%%%%%%%%%%%%%%%%%%%%%%%%%%%%%%%%%%%%%%%%%%%%%
%REMARK Figures for different radius and charges
%%%%%%%%%%%%%%%%%%%%%%%%%%%%%%%%%%%%%%%%%%%%%%%%%%%%%%%%%%%%%%%%%%%%%%%%%%%%%%%%%%%%%%
\begin{figure}
\begin{center}
\begin{tabular}{cc}
\includegraphics[width=0.5\textwidth]{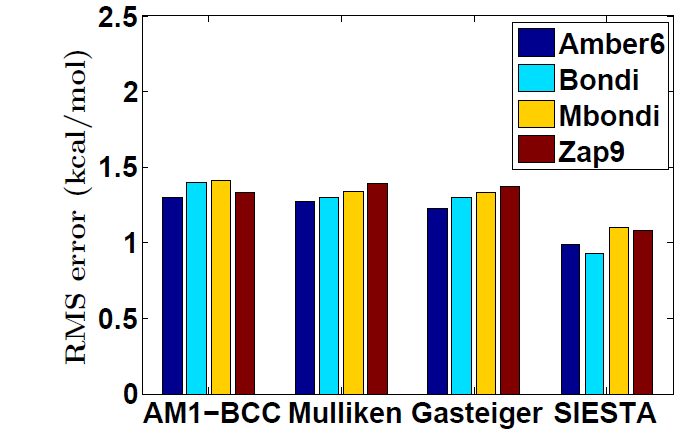}
\includegraphics[width=0.5\textwidth]{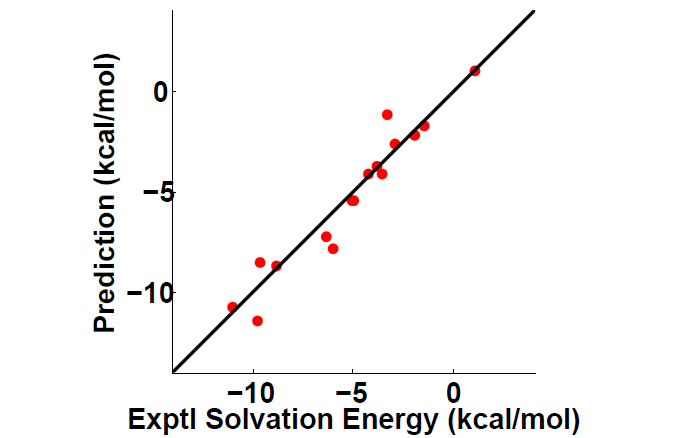}
\end{tabular}
\end{center}
\caption{The prediction results for the SAMPL0 blind test set, the left chart shows the RMS errors between the experimental and prediction solvation free energies. The right chart shows the optimal predictions of the solvation free energies for the SAMPL0 test set.}
\label{SAMPL0-figs}
\end{figure}

\begin{figure}
\begin{center}
\begin{tabular}{cc}
\includegraphics[width=0.5\textwidth]{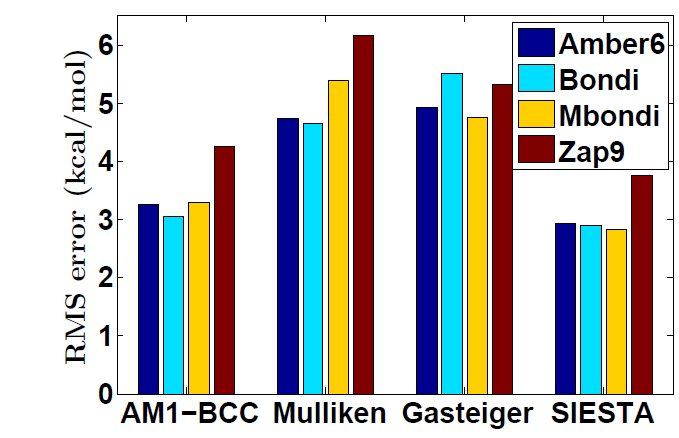}
\includegraphics[width=0.5\textwidth]{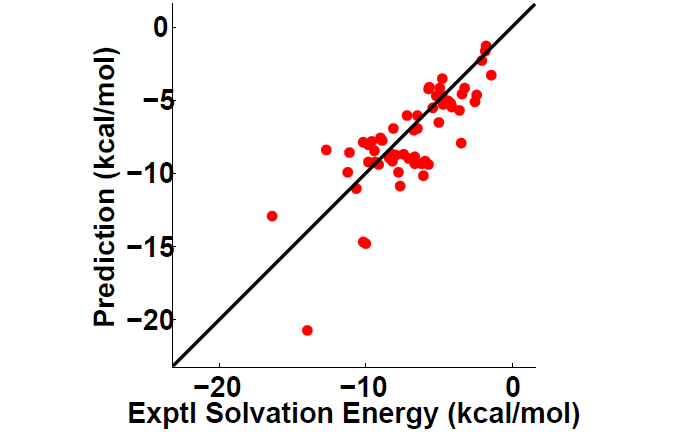}
\end{tabular}
\end{center}
\caption{The prediction results for the SAMPL1 blind test set, the left chart shows the RMS errors between the experimental and prediction solvation free energies. The right chart shows the optimal predictions of the solvation free energies for the SAMPL1 test set.}
\label{SAMPL1-figs}
\end{figure}

\begin{figure}
\begin{center}
\begin{tabular}{cc}
\includegraphics[width=0.5\textwidth]{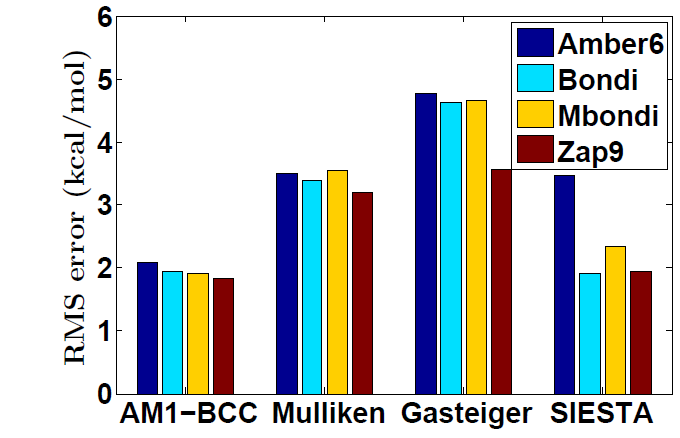}
\includegraphics[width=0.5\textwidth]{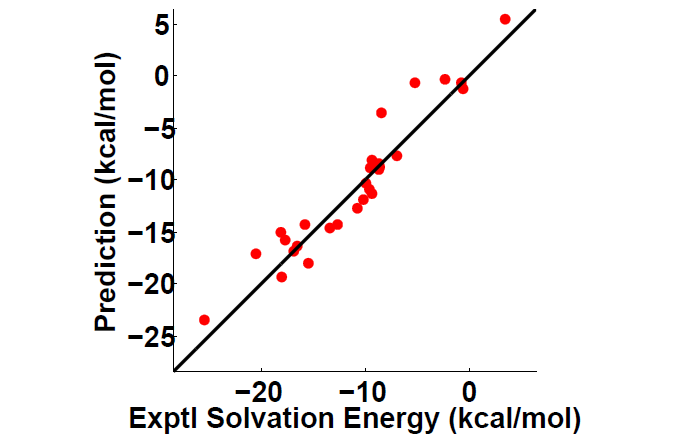}
\end{tabular}
\end{center}
\caption{The prediction results for the SAMPL2 blind test set, the left chart shows the RMS errors between the experimental and prediction solvation free energies. The right chart shows the optimal predictions of the solvation free energies for the SAMPL2 test set.}
\label{SAMPL2-figs}
\end{figure}

\begin{figure}
\begin{center}
\begin{tabular}{cc}
\includegraphics[width=0.5\textwidth]{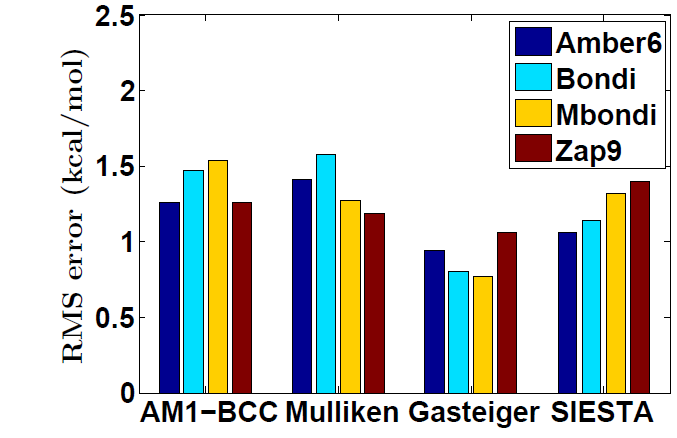}
\includegraphics[width=0.5\textwidth]{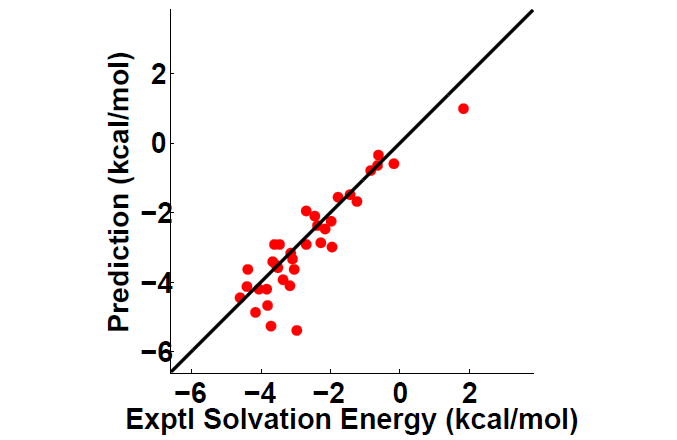}
\end{tabular}
\end{center}
\caption{The prediction results for the SAMPL3 blind test set, the left chart shows the RMS errors between the experimental and prediction solvation free energies. The right chart shows the optimal predictions of the solvation free energies for the SAMPL3 test set.}
\label{SAMPL3-figs}
\end{figure}

\begin{figure}
\begin{center}
\begin{tabular}{cc}
\includegraphics[width=0.5\textwidth]{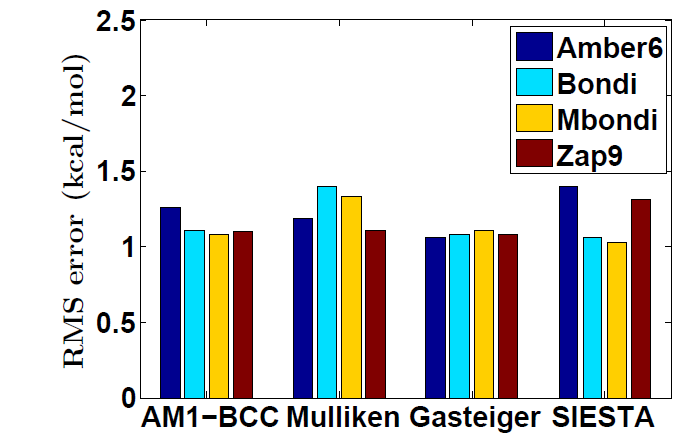}
\includegraphics[width=0.5\textwidth]{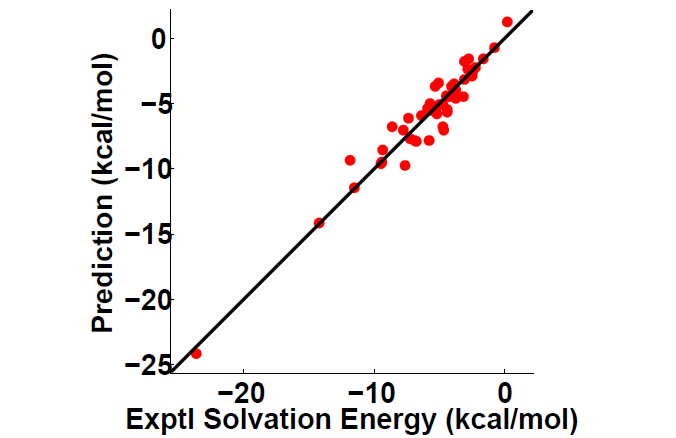}
\end{tabular}
\end{center}
\caption{The prediction results for the SAMPL4 blind test set, the left chart shows the RMS errors between the experimental and prediction solvation free energies. The right chart shows the optimal predictions of the solvation free energies for the SAMPL4 test set.}
\label{SAMPL4-figs}
\end{figure}

\paragraph{SAMPLx blind predictions}
In this section, we present results of the blind solvation free energy predictions based on the  proposed protocol. All the  SAMPL0-SAMPL4 challenges for solvation free energies are considered. We predict the solvation free energies with a leave-SAMPLx-out approach, in which the SAMPLx molecules and their experimental solvation free energies are regarded as unknown while information of other molecules is utilized to predict selected SAMPL test set, based on molecular formulas.

We first consider SAMPL0 test set, in which molcules are diverse. One molecule in this test set contains Br atom, for which our polarizable PB model does not an appropriate pseudo-potential. The structures of the SAMPL0 molecules can be found in the literature \cite{Nicholls:2008solvation}.  Tables S1-S4 \cite{Supporting} present the solvation free energies for the SAMPL0 test set obtained from both experiments and blind predictions using different charges and atomic radii parametrizations. The left chart of Fig. \ref{SAMPL0-figs} shows the plot of  RMS  errors for each charge and atomic radius combination. It is clear that the change in atomic radii  has a minor influence on the accuracy of  predictions, while the change in charge force fields has a much more significant influence. In other words, the solvation free energy prediction for this test set is more sensitive to the charge parametrization than the atomic radii parametrization. In general, our KSDFT based polarizable Poisson model provides better  predictions than those of other  charge force fields. It is noted that the SIASTA based polarizable Poisson model with Amber Bondi radius parameters delivers the best solvation free energy prediction.  This result is depicted in the right chart of Fig. \ref{SAMPL0-figs}. The associated RMS error (0.93 kcal/mol)  appears to be better than that  in the literature \cite{Nicholls:2008solvation} (i.e., 1.71 kcal/mol for the {\it full} SAMPL0 test set of 17 molecules). The best prediction in the literature has an RMS error of 1.34 kcal/mol \cite{Kehoe:2012}. Even though for the SIESTA based {\it Ab Initio} charge calculation, the molecule contains the Br atom is neglected, based on the general pattern that when any other force field used, the prediction for that omitted molecules is very well. We believe with proper pseudopotential for Br atom, our polarizable Poisson model can deliver an accurate prediction for this molecule also. Moreover, for this test set, all the other three charge parametrizations can provide a similar level of solvation free energy prediction.

%Table of RMS error
\begin{table}
\centering
\caption{The RMS errors of the solvation free energy prediction with different methods. The RMS errors inside and outside the parenthesis denote for the prediction errors include and exclude the molecules contains Br atom. All with unit kcal/mol}
\begin{tabular}{ c|l|cccc }
%\hline
%\multicolumn{6}{ c }{The RMS error of the solvation free energy prediction with different method} \\
\hline
Test set                & Radius        &BCC          & Mulliken    & Gasteiger  &SIESTA \\ \hline
\multirow{4}{*}{SAMPL0} &Amber 6        &1.30 (1.26)  &1.27 (1.25)  &1.23 ({\bf 1.20}) &0.99 (NA) \\
                        &Amber Bondi    &1.40 (1.37)  &1.30 (1.27)  &1.30 (1.27) &{\bf 0.93} (NA) \\
                        &Amber Mbondi2  &1.41 (1.37)  &1.34 (1.32)  &1.33 (1.29) &1.10 (NA) \\
                        &ZAP9           &1.33 (1.29)  &1.39 (1.37)  &1.37 (1.33) &1.08 (NA) \\
\hline
\multirow{4}{*}{SAMPL1} &Amber 6        &3.26 (3.27)  &4.74 (4.77)  &4.92 (4.96) &2.92 (NA) \\
                        &Amber Bondi    &3.06 ({\bf3.07})  &4.65 (4.68)  &5.52 (5.55) &2.89 (NA) \\
                        &Amber Mbondi2  &3.29 (3.30)  &5.39 (5.41)  &4.76 (4.82) &{\bf 2.82} (NA) \\
                        &ZAP9           &4.26 (4.35)  &6.16 (6.16)  &5.33 (5.45) &3.76 (NA) \\
\hline
\multirow{4}{*}{SAMPL2} &Amber 6        &2.09 (2.11)  &3.51 (3.59)  &4.78 (4.86)  &3.46 (NA) \\
                        &Amber Bondi    &1.95 (1.97)  &3.38 (3.47)  &4.62 (4.72)  &{\bf 1.90} (NA) \\
                        &Amber Mbondi2  &{\bf 1.90} ({\bf 1.96})  &3.55 (3.66)  &4.65 (4.76)  &2.35 (NA) \\
                        &ZAP9           &2.05 (2.03)  &3.19 (3.15)  &4.56 (4.51)  &1.93 (NA) \\
\hline
\multirow{4}{*}{SAMPL3} &Amber 6        &1.28  &1.42  &0.97  &1.08 \\
                        &Amber Bondi    &1.47  &1.58  &0.82  &1.16 \\
                        &Amber Mbondi2  &1.47  &1.58  &0.82  &1.16 \\
                        &ZAP9           &1.55  &1.28  &{\bf 0.78}  &1.33 \\
\hline
\multirow{4}{*}{SAMPL4} &Amber 6        &1.28  &1.20  &1.08  &1.41 \\
                        &Amber Bondi    &1.12  &1.41  &1.10  &  1.07  \\
                        &Amber Mbondi2  &1.09  &1.33  &{\bf 1.03}  &1.04 \\
                        &ZAP9           &1.12  &1.12  &1.09  &1.32 \\
\hline
\end{tabular}
\label{RMSerrorlist}
\end{table}

%The RMS errors of the solvation free energy prediction for SAMPL1 test set with different parametrization are illustrated the left chart of Fig. \ref{SAMPL1-figs}.  Obviously SAMPL1 test set is relative large  and involves very different molecules. Except for one case, RMS errors for all predictions are  above 2.00kcal/mol, which indicates the difficulty of making a good prediction for this test set. Our best blind prediction, obtained with ZAP9 radius and SIESTA based polarizable Poisson model, has an RMS error of 1.88kcal/mol. This best prediction is illustrated in the right chart of Fig. \ref{SAMPL1-figs}.
%In a comparison, the best result in the literature, to our best knowledge, has the RMS error of xxx cite{}.

We next consider the SAMPL1 challenge set for  solvation predictions  \cite{SAMPL1:2009}. This set contains not only a largest number of 63 molecules, which is the largest  among all the Sampl test sets, but also many molecules with extremely complicated structures. The detail description of this set is given in the literature \cite{SAMPL1:2009}. Most molecules in this set are druggable and very complex. The difficulty of  this test set comes from two aspects: on the one hand, the structures of  molecules in this set are very complicated. On the other hand, the knowledge in the training set that can be used for predicting the solvation free energies of these molecules is rare or insufficient. In addition to  the molecular complexity, the reported experimental solvation free energies also admit large uncertainty \cite{SAMPL1:2009}. Most earlier computational predictions report RMS errors of 3 to 4 kcal/mol when some extremely complex molecules are excluded. Some best prediction for the whole set has an RMS error of 2.45 kcal/mol \cite{Kehoe:2012}.
The best performance was shown to give an RMS error of 2.4 kcal/mol on a subset of the SAMPL1 test set that contains only 56 molecules \cite{Truhlar:2009SAMPL1}. Tables S5-S8  \cite{Supporting} present the solvation free energies for the SAMPL1 test set obtained from both experiments and blind predictions using different charges and atomic radii. Figure \ref{SAMPL1-figs} plots the present blind predictions.  One molecule (bromacil) that contains a Br atom is considered by all the  charge models except for the polarizable PB method. It is noted that two of present approaches, namely, the AM1-BCC semi-empirical charges and the \textit{Ab Initio} charge provide relatively accurate predictions for this test set.
When Amber Mondi2 atomic radii together with the SIESTA charge calculation applied, the optimal solvation free energy prediction is achieved with RMS error 2.82 kcal/mol. When BCC charges and  the Amber Bondi radii are used, the prediction for the whole SAMPL1 set without a molecule that contains Br atom has an RMS error of 3.06 kcal/mol, the RMS error is 3.07 kcal/mol for the same test with all molecules. For this test set, the large prediction RMS error mainly due to the extremely large error in predicting solvation free energies for some complex molecules, for which the RMS error can be as large as 15 kcal/mol. This  unreasonable prediction is due to the fact that inappropriate molecule force field parametrization yields unreasonable  electrostatic solvation free energies, which in turn leads to erroneous solvation free energy prediction. Similar to the  SAMPL0 test set, the prediction is more sensitive to the charge parametrization.  Nevertheless, the atomic radius parametrization also plays an critical role in the prediction accuracy for this test set.

%While for the Gasteiger and Mulliken charges, the prediction results are basically the same as those in the literature \cite{SAMPL1:2009,Kehoe:2012}.
%In all the predictions, the molecule that contain Br atoms is not included.

SAMPL2 is another difficult test set with almost the same level of difficulty as the SAMPL1 test set \cite{David:2010solvation}. Compared with  SAMPL1 test set it contains a few complex molecules, and most molecules in this test set are drug like ones as well. Contrary to  molecules in  SAMPL1 test set, this set has less uncertainty in the experimental solvation free energies.   As shown in  Tables S9-S12  \cite{Supporting}, the experimental solvation free energies of this test set distribute over a wide range. Using all-atom molecular dynamics simulations and multiple starting conformations for blind prediction, Klimovich and Mobley reported an RMS error of 2.82 kcal/mol over the whole set and 1.86 kcal/mol over all the molecules except several hydroxyl-rich compounds \cite{David:2010solvation}. Some best prediction has an RMS error of 1.59 kcal/mol \cite{Kehoe:2012}. In the present work, the molecule containing an I atom (5-iodouracil) is excluded in all calculations. Additionally,   5-bromouracil has a Br atom is excluded in the polarizable PB model.
 Tables S9-S12  \cite{Supporting} %\ref{SAMPL2-amber6}-\ref{SAMPL2-ZAP9},
list the details of the present predictions. The RMS errors from various radius and charge force fields are given Fig. \ref{SAMPL2-figs}.  Apparently, this test set has a strong force field dependence as well. The RMS errors vary very much from one charge force field to another. However, the performance of these predictions has a weak radius dependence. The best prediction, obtained from a combination of Amber MBondi2 radius parameters and BCC charge force field, or Amber Bondi radius together with modified SIESTA charge, both have the RMS error of 1.90 kcal/mol when the molecule with a Br atom is excluded in the prediction. When all molecules are included, the combination of Amber MBondi2 radius and BCC charge parametrization gives the optimal prediction of  RMS error 1.96 kcal/mol as depicted in the right chart of Fig. \ref{SAMPL2-figs}. Compared to the prediction of SAMPL1 test set, these predictions are more accurate, due to two reasons. First, the molecules in this set are slightly simpler and experimental uncertainly is less severe for the deterministic prediction. Second, in our  parametrization of  the nonpolar solvation free energy prediction, we have more similar molecules in training set, which enables us to obtain better nonpolar solvation free energy parameters in the nearest neighbor based approach. In contrast, in the SAMPL1 prediction,  when the nearest neighbor approach is applied, the nearest molecules selected from the training  set differ much  from SAMPL1 molecules.

%In a comparison, the best result in the literature, to our best knowledge, has the RMS error of xxx cite{}.

%Can provide much better results not only due to the slightly simple set itself, but also due to the knowledge based models, that we have more accurate knowledge to learn the
%parameters in the nearest neighbor based models. Compare to the test in S1, in which when all the S1 moleucles are left out, when nearest neighbor approach used, the
%nearest molecules detected already has a huge difference from the original molecule
%Mul and Gas seems not a good charge parametrization force field for complex molecules based on S1 and S2 sets.

SAMPL3 test set is a relatively easy one with 36 solute molecules \cite{SAMPL3:2012}. Its  molecular structures are less versatile than the earlier  test sets. Additionally, in this test set, there is no molecule that involves Br or I atom. One of difficulties in the prediction of this set is the lack of similar molecules  in our database when all the SAMPL3 molecules are left out. The best prediction in the literature offers  an RMS error of 1.29 kcal/mol \cite{Kehoe:2012}.   Our blind predictions and experimental results for SAMPL3 test set are given in Tables S13-S16  \cite{Supporting}. %\ref{SAMPL3-amber6}-\ref{SAMPL3-Zap9}.
The RMS errors of our blind predictions for the SAMPL3 test set are plotted in Fig. \ref{SAMPL3-figs}. In this case, the accuracy of predictions also depends strongly on  charge force fields and weakly on radius parameters.  In general, Gasteiger charges are superb for this test set and their RMS errors are always smaller than 1 kcal/mol. Our best prediction, obtained from the combination of ZAP9 radius parameters and Gasteiger charge force field, has a small RMS error of 0.78 kcal/mol and is depicted in the right chart of Fig. \ref{SAMPL3-figs}. In general Gasteiger charge performs very poor in the solvation free energy prediction for  previous tests on complex molecules,  while  we see that this charge parametrization method is superior for chlorinated hydrocarbon molecules in the present test set. Unfortunately, there is no uniformly optimal parametrization for all the molecules. This fact   motivates us to seek a solvation free energy prediction that does not heavily depends on the force fields parametrization.

%In a comparison, the best result in the literature, to our best knowledge, has the RMS error of xxx cite{}.

SAMPL4 test set is studied by using the proposed methods as well.  A key feature of this test set is that its molecules are diversified.   As shown in Tables S17-S20  \cite{Supporting}, this test set also involves a wide range of solvation free energies.
%\ref{SAMPL4-amber6}-\ref{SAMPL4-Zap9},
However, the structures of these molecules are not as complex as those in both SAMPL1 and SAMPL2 test sets, which indicates a slightly easier task for solvation free  energy prediction. This test set was studied by a number of researchers in the literature and the best prediction in the literature  has the RMS error of 1.2 kcal/mol  \cite{David:2014solvation}. Figure \ref{SAMPL4-figs} illustrates the RMS errors of our predictions for a total of 16 charge and radius combinations. Compared with those in the literature, all of our predictions are of high quality. Our best result has the RMS error of 1.03 kcal/mol and the corresponding results are given in the right chart of Fig. \ref{SAMPL4-figs}.

Table \ref{RMSerrorlist} lists the RMS errors of our blind predictions of solvation free energies for all five test sets using a total of 16 charge and radius implementations. Some comments are in order. First,  all predictions are quite sensitive to  charge force fields and relatively less dependent on radius parameter selection. Additionally, it is difficult to identify a clear winner over all the test sets ---  some approaches perform better in one or two test sets, but do not do well in the rest test sets. This phenomenon  highlights the difficulty of designing optimal models for solvation analysis. Moreover, the KSDFT based polarizable Poisson model is more often to provide better predictions over most charge force fields and radius parameters. This indicates a potential to develop better physical models by improving the quantum charge density calculation.  Finally, we point out that the blind prediction results presented in the present work are the state-of-the-art compared to those in the literature  \cite{Nicholls:2008solvation, David:2014solvation, Guthrie:2014solvation, Klamt:2014solvation, David:2012solvation, David:2009solvation,  David:2006JCTCsolvation, David:2008JPCBsolvation, Wujianzhong:2014, SAMPL1:2009, SAMPL:2010, SAMPL3:2012,Kehoe:2012}.

%\begin{table}[!ht]
%\centering
%\caption{\footnotesize{The experimental and predicted solvation free energy for the SAMPL 1 test set. $\Delta G_{exp}$ is the experimental solvation free energy, $\Delta G^{pred}_{AM1-BCC}$, $\Delta G^{pred}_{Mulliken}$, $\Delta G^{pred}_{Gasteiger}$, and $\Delta G^{pred}_{SIESTA}$ are the predicted solvation free energy with four different types of charges. The Amber6 ZAP9 force field are employed for the atomic radius parameterization. The energy data are all in the unit of $kcal/mol$.}}
%\begin{tabular}{llllll}
%%{@{\extracolsep{\fill}}llllll@{}}
%\cline{1-6}
%\footnotesize{Solute Name} &\footnotesize{$\Delta G^{exp}$}  &\footnotesize{$\Delta G^{pred}_{AM1-BCC}$}  &\footnotesize{$\Delta G^{pred}_{Mulliken}$} & \footnotesize{$\Delta G^{pred}_{Gasteiger}$} & \footnotesize{$\Delta G^{pred}_{SIESTA}$}  \\
%\hline

\section{Concluding Remarks}\label{conclusion}

In this work, we propose a   protocol to parametrize implicit solvent models for the blind prediction of  solvation free energies. The proposed protocol parametrizes nonpolar solvation model by scoring and ranking algorithms while utilizes either the Poisson model or a density functional theory (DFT) based polarizable Poisson model for polar solvation free energy calculations.  For the parametrization of  nonpolar solvation models, we first utilized the assumption that molecules with the same functional group admit the same parametrization of the nonpolar solvation energy functional. For complex poly-functional-group molecules, we develop a scoring procedure to determine the optimal relative weight of each functional group. For extremely complex molecules that fails the functional group scoring method, we further develop a molecule ranking algorithm to select an optimal set of nearest neighbor molecules for parameter training. We construct atomic features for the  molecule ranking. Finally, we systematically integrate the above mentioned models and algorithms into a robust protocol for blind solvation free energy prediction.

In the present work, we considered an experimental database of 668 solvation molecules, see Table S21 \cite{Supporting}, the largest database ever constructed for solvation, to validate our approach. Among them, SAMPL0 to SAMPL4 test sets are paid special attention. For the Poisson model or DFT based  polarizable Poisson  model,  four sets of atomic radius parameters (i.e., Amber 6, Amber bondi, Amber mbondi and ZAP9 radii) are combined with four sets of charge force fields (i.e., AM1-BCC, Mulliken, Gasteiger and SIESTA DFT) to arrive at a total of 16 different implementations. The resulting polar solvation free energies are utilized in our parametrization for blind predictions.   We first carry out the leave-one-out validation of the whole database.  The AM1-BCC charge force field delivers  a low  root mean square (RMS) error of 1.33 kcal/mol, which is the lowest for such a large test database, to our best knowledge. We further conduct a series of leave-SAMPLx-out blind tests. On average, the BCC parametrization in the Poisson model and DFT based polarizable Poisson model performs better than other charge force fields, especially for predicting the solvation free energies of the complex molecules. We obtain some of best known results.  The optimal RMS errors for SAMPL0-SAMPL4 are respectively,  0.93, 2.82, 1.90, 0.78, and 1.03 kcal/mol, which again, are some of the best to our best knowledge.

From the solvation free energy predictions, particularly on SAMPL1 and SAMPL2 test sets, we conclude that atomic charge parametrization is extremely important for the present physical models, namely, the Poisson model or the KSDFT based  polarizable Poisson  model. Without an appropriate charge parametrization, the prediction errors can be  amplified  for molecules with complex structures. In general, for four charge parametrization methods, both the semi-empirical BCC charge and the {\it Ab Initio} charge calculation from the generalized KSDFT can provide relatively reliable charge assignments. For this reason, a solvation free energy prediction method that does not heavily depend on the molecule parametrization is under our consideration. Essential idea is that if one does not partition the solvation free energy into two isolated parts, the prdeiction errors in the polar solvation free energy will not propagate to the nonpolar solvation free energy prediction. Alternative, our differential geometry based solvation model that dynamically couple polar and nonpolar models \cite{Wei:2009,ZhanChen:2010a} might also provide a less sensitive approach.

This work can be improved in a number of ways. First, the classification  of the database into functional groups is not unique.  Future study will explore optimal molecules partition. Additionally, the selection and computation of atomic features need to be further investigated.  It is possible to construct an optimal set of atomic features for solvation analysis and prediction. Further, in the current work, molecular ranking for nearest neighbor searching is not optimal yet. More sophisticated machine learning and/or deep learning algorithms can be developed for this purpose.  Finally, a more versatile DFT solvers can be utilized to further improve our in-house  polarizable Poisson model. These issues are under our consideration.

\section*{Acknowledgments}
This work was supported in part by NSF Grant   IIS- 1302285 and MSU Center for
Mathematical Molecular Biosciences Initiative.
BW thanks Dr. Pengfei Li, Professor Ray Luo, Professor David Mobeley, and Professor Wei Yang for helpful discussions.
GWW acknowledges Nathan Baker,  Michael Gilson  and Weitao Yang for useful discussions.

\vspace{1cm}
%\bibliographystyle{abbrv}
%%\bibliographystyle{plain}
%\bibliography{refs}

\end{document}